\documentclass[prd,nofootinbib,longbibliography,twocolumn]{revtex4-1}

\usepackage[mathscr]{eucal}
\usepackage[latin1]{inputenc}
\usepackage{amsmath}
\usepackage{amsfonts}
\usepackage{amssymb}
\usepackage[usenames,dvipsnames]{color}
\usepackage[pdftex]{graphicx}
\usepackage[raggedleft]{sidecap}

\usepackage{subfig}
\usepackage{graphicx}
\usepackage{sidecap}
\usepackage{hyperref}

\usepackage{enumerate}

\usepackage{colordvi}
\newcounter{mnotecount}[section]

\newcommand{\comment}[1]{}
\usepackage{enumerate}

\newcommand{\be}{\nopagebreak[3]\begin{equation}}
\newcommand{\ee}{\end{equation}}
\newcommand{\ba}{\nopagebreak[3]\begin{eqnarray}}
\newcommand{\ea}{\end{eqnarray}}

\definecolor{celeste}{rgb}{.23,.5,.7}
\newcommand{\Feyn}[1]{#1\kern-0.65em/}
\def\be{\begin{eqnarray}}
\def\ee{\end{eqnarray}}
\def\bc{\begin{center}}
\def\ec{\end{center}}

\begin{document}
 \title{Relative information at the foundation of physics\\
 {\rm \em \small Second prize in the 2013 FQXi context ``It From Bit or Bit From It?"} 
 }
\date{\today}

    \author{Carlo Rovelli}
\affiliation{CPT, CNRS UMR7332, Aix-Marseille Universit\'e and Universit\'e de Toulon, F-13288 Marseille, EU}


\begin{abstract}                
\noindent 
Shannon's notion of relative information between two physical systems can  function as foundation for statistical mechanics and quantum mechanics, without referring to  subjectivism or idealism. It can also represent a key missing element in the foundation of the naturalistic picture of the world, providing the conceptual tool for dealing with its apparent limitations. I comment on the relation between these ideas and Democritus. 

\end{abstract}

\maketitle

\section{Is there a subjective element in statistical mechanics?}

Thermodynamical quantities such as entropy and temperature depend on the macroscopical variables chosen to describe systems with many degrees of freedom. They depend on coarse-graining.  For instance, entropy can be defined (in the microcanonical) in terms of the the number of microstates compatible with what we know about the system.  With this definition, it changes if we know more.  This appears to insert a puzzling subjective element in physics. There is a tension with the fact that termodynamical laws seem to hold quite independently on any choice by us.  Is the Sun ``hot" just because we ``choose" a certain coarse graining for describing it? Does entropy increases because of our choices?

The way out of the puzzle is simple. Entropy is neither something inherent to the microstate of a system, nor something depending on our subjective ``knowledge" about it.  Rather, it is a property of certain (macroscopic) variables.  For instance, the full state of a gas in a box is described by the position and velocity of its molecules. No entropy so far.  But volume, total energy and (time averaged) pressure on the box boundaries are well defined functions of this state, and Entropy is a function of \emph{these}.  This is the first step. 

Now consider a situation where the gas interacts with a \emph{second} system coupled only to  volume, total energy and pressure of the gas (for instance, it interacts with the gas by a thermometer and a spring holding a piston). Then the physical interactions between the gas \emph{and this system} are \emph{objectively} described by thermodynamics.  

In other words, it is not an arbitrary or subjective choice of a coarse-graining that makes thermodynamics physically relevant: it is the concrete way another physical system is coupled to the gas. If the coupling is such that it depends only on certain gas macroscopic variables, then the physical interactions of the gas and this system are \emph{objectively} well governed by thermodynamics.

This key observation clarifies the role that information plays in physics. Entropy, indeed, is information: in the micro-canonical language entropy is determined by the number of microstates compatible with a given macrostate.  The number of states in which something can be, is precisely the definition of ``information" (more precisely, ``lack of information") given by Shannon in his celebrated 1948 work that started the development of information theory \cite{Shannon:1948fk}.  But ``information", that is, the number of alternatives compatible with what we know, is not significative in physics insofar as it depends on idealistic subjective knowledge: it is relevant in physics when it refers to the \emph{interaction} between two systems where the effects of the interaction on the second  depend only on few variables of the first, and are independent on the rest of the variables. Under these circumstances, the number of states of the first system which are not distinguished by these variables is the number of Shannon ``alternatives", relevant for the definition of thermodynamical entropy.  Here ``information", counts the number of states of a system which behave equally in the interaction with a second system.   

Therefore the information relevant in physics is always the \emph{relative} information between \emph{two} systems.  There is no subjective element in it: it is fully determined by the state \emph{and} the interaction Hamiltonian which dictates which variables are the relevant ones in the interaction.  

Pictorially: it is not the microstate of the Sun which is hot, it is the manner the Sun affects the Earth which is \emph{objectively} hot.

\section{Relative irreversibility}

Reconsider the quintessential irreversible phenomenon: a cup falls to the floor and breaks, in the light of the observation above. On one account this is obviously an irreversible phenomenon, but is it so on any possible account?  The event is one among the many possible dynamical evolutions of a bunch of molecules. What makes the starting configuration more ``special" that the final one?  Something does so, but it is not in the microstate of the molecules: it is the manner we describe, or better, at the light of the previous section, we interact with it. It is because of our  macroscopic account of the cup, dictated by the variables we interact with, that the initial state is special and therefore entropy increases.

To illustrate this, consider a box full of balls, characterized by two properties, say color and electrical charge.  Say they have two possible colors:   white and  black; and two possible value of the charge:    neutral and  charged. Consider a microstate $Col$ where  white balls are on the left of the box and   black balls on the right, while charge is randomly distributed. And consider a different microstate  $Ch$, where  charged balls are on the left and   neutral balls on the right, while color is randomly distributed. To normal eyes, $Col$ looks as a low-entropy state and $Ch$ as a high entropy state.   But to a person who is color blind but has an electrometer it us $Ch$ that looks low-entropy and $Col$ that appears to have high entropy.  Who is right? Both, of course. Entropy is relational: it pertain to the relation between two interacting systems, not to a single system. 

Could the breaking cup be observed by somebody else, coupling differently to it, as a process where entropy decreases? Yes of course. Imagine each fragment of the cup moving to a picture of itself on the ground, pictured in color to which you are color blind. 

If these considerations are correct, then the irreversibility of the worlds is to be understood as a property of the couplings between systems, rather than a property of isolated systems. 

\section{The limits of microphyscs without information}

The idea that the world can be described as a vast see of interacting atoms and nothing else, can be traced back to the ancient atomism of Democritus. The naturalistic and materialistic world view of Democritus was soon criticized by Plato and Aristotle on the ground that it fails to account for the forms, or the objects, that we see in the world.  What makes a certain ensemble of atoms into a given object we recognize? Plato and Aristotle (in different manners) wanted to add ``forms" to the naturalistic view of Democritus.   For Plato, a horse is not just an aggregate of matter: it is an imprecise realization of the abstract form (``idea") of a horse. For Aristotle the same horse is the union of its substance and its form.  But if the form is something above the substance, what is it? 

What is it that makes a random disposition of molecules into a cup?  Which the properties of the Democritean atoms generate collective variables? And how?

In fact, Democritus's idea was more subtle than everything being just atoms. Democritus says that three features are relevant about the atoms: the shape of each individual atom, the order in which they are disposed, and their orientation in the structure. And Democritus uses then a powerful metaphor: like twenty letters of an alphabet can be combined in innumerable manners to give rise to comedies or tragedies, similarly the atoms can be combined in innumerable manners to give rise to the innumerable phenomena of the world.

But what is the relevance of the way in which atoms combine, in a world in which there is nothing else than atoms?  If they are like letters of an alphabet, whom do they tell stories to? 

I think that the key of the answer brings us back to the observation in the first section: physical systems interact and affect one another. In the course of these interactions, the way one  happens to be, leaves traces on the way another is: correlations are established. 

Following Shannon, we can say that a system $s$ has information about a system $S$ if there is a physical constraint such that the number of total states of the two systems is smaller than the product of the number of states of each. For instance: if the system $s$ can be in the states $a$ and $b$ and the system $S$ can be in the states $A$ and $B$, but there is a physical constraint (say do to the way the two have interacted) that forbids the combinations $(a,B)$ and $(b,A)$, thus allowing only the two states  $(a,A)$ and $(b,B)$, then we say that $s$ has (one bit of) information about $S$.  In words, if we see the state of $s$, we also know the state of $S$.   Physical interactions determine constraints among systems: if a tree happen to fall on my head, then I cannot be standing smiling anymore: I have some information about the tree. 

Thus, systems have necessarily information about one another, in the sense of Shannon.  The lack of information that a system has about another is precisely the entropy of the second with respect to the first. It is relevant for the interactions with the first. It is the conventional thermodynamical entropy. 

Before pursuing this line of thinking, let me bring quantum theory into the picture. 

\section{Quantum theory}

The discovery of quantum theory has sharpened the role of information in our understanding of the world.  If we measure the state of a system with a certain precision, the resulting information  specifies a region $R$ of the phase space of the system.  The unit of phase space volume is action (length${}^{2}$ x mass x time${}^{-1}$) per degree of freedom. In classical mechanics we can in principle arbitrarily refine measurements, therefore there is always a continuous (infinite) amount of missing information about a system, whatever the precision of the measurement.

No longer so after the discovery of quantum theory.  If we measure the energy of a harmonic oscillator and we obtain the result that this is between $E_1$ and $E_2$, then there is only a \emph{finite} number of possible values that the energy can have.  This is given by the area of the region of phase space included between the two surfaces $E_1$ and $E_2$, divided by the Planck constant.  

This is a general result: for all quantum systems, there is only a finite number of the orthogonal states per each finite region of phase space.  The Planck constant determines the minimal phase space volume.  Phase space volume measures the (missing) information we have about a system. It follows that quantum mechanics affirms that information is no longer continuous as in classical physics. It is discrete, and the Planck constant is the minimal unit of information. 

This leads to a first principle at the basis of quantum theory:  \emph{The information contained in any finite region of the phase space of any system is finite.}  

This principle does not exhaust quantum theory, because it holds for any discrete classical system as well. What further characterises quantum theory is that information can become ``irrelevant", and be renewed. By this I mean the following. If we have measured a system, the information we have about it allows us to predict its future. In quantum theory, we can always add \emph{new} information to the state of a system, even after we have reached maximal information about it. By doing so, part of the old information becomes irrelevant. That is, is has no effect on future predictions.  The typical case is a sequence of measurements of spins along different axes, in a two-state system. Each measurement brings novel information and makes the previous one irrelevant. 

This leads to the second principle at the basis of quantum theory:  \emph{It is always possible to acquire new information about a system.}  

The combination of these two principles generates the entire mathematical structure of quantum theory, up to some technical aspects, as was shown in \cite{Rovelli:1995fv}.  Thus, \emph{relative information} that systems have about one another is a key language for grounding quantum theory. 

Let us now remember that spacetime geometry is dynamical.  Then any physical system must include the its spacetime region. This implies that there is natural identification between systems and spacetime regions.  The interactions between spacetime regions are quantum interactions between systems. Therefore there is an exchange of informations across spacial regions. These are quantized and discrete, because information is discrete.  The quantum discreetness, combined to the fact that geometry is dynamical and therefore quantized, leads to the discretization of space, idea that can be traced back to the thirties \cite{Bronstein:1936kx,Bronstein:9vn} and has been concretized more recently. Discreetness translates into the discreetness of the area of two-dimensional surfaces \cite{Rovelli:1994ge,Ashtekar:1996eg}. The discreteness of the area is an immediate reflex of the discreetness of the quantum information that can be transmitted across these surfaces. 

\section{Reality and information}

It seems to me that this ensemble of considerations conspire towards a picture where the fog begins a bit to dissipate over the intriguing role of information at the foundation of physics. 

Information that physical systems have about one another, in the sense of Shannon, is ubiquitous in the universe. It has the consequence that on top of the microstate of a system we have also the informational state that a second system $O$ has about any system $S$.  

The universe is not just the position of all its Democritean atoms.  It is also the net of information that all systems have about one another.  Objects are not just aggregate of atoms. They are configurations of atoms singled out because of the manner a given other system interacts with them. An object is only such with respect to an observer interacting with it. 

Among all systems, living ones are those that selection has led to persist and reproduce by, in particular, making use of the information they have about the exterior world. This is why we can understand them in terms of finality and intentionality. They are those that have persisted thanks to the finality in their structure. Thus, it is not finality that drives structure, but selected structures define finality. Since the interaction with the world is described by information, it is by dealing with information that these systems most effectively persist.  This is why we have DNA code, immune systems, sensory organs, neural systems, memory, complex brains, language, books, MAC's and the ArXives.  To maximize the management of information. 

The statue that Aristotle wants made of more than atoms, is made by more than atoms: it is something that pertains to the interaction between the stone and brain of Aristotle, or ours. It is something that pertains to the stone, the goddess represented, Phidias, a woman he met, our education, and else. The atoms of that statue talk to us precisely in the same manner in which a white ball in my hand ``says" that the ball in your hand is also white, if the two are correlated.  By carrying information. 

This is why, I think, from the basis of genetics, to the foundation of quantum mechanics and thermodynamics, all the way to sociology and quantum gravity, the notion of information has a pervasive and unifying role. The world is not just a blind wind of atoms, or general covariant quantum fields.  It is also the infinite game of mirrors reflecting one another formed the correlations among the structures made by the elementary objects.   To go back to Democritus metaphor: atoms are like an alphabet, but an immense alphabet so rich to be capable of reading itself and thinking itself.  In Democritus words:

``The Universe is change, life is opinion that adapt itself". 


\begin{thebibliography}{6}%
\makeatletter
\providecommand \@ifxundefined [1]{%
 \@ifx{#1\undefined}
}%
\providecommand \@ifnum [1]{%
 \ifnum #1\expandafter \@firstoftwo
 \else \expandafter \@secondoftwo
 \fi
}%
\providecommand \@ifx [1]{%
 \ifx #1\expandafter \@firstoftwo
 \else \expandafter \@secondoftwo
 \fi
}%
\providecommand \natexlab [1]{#1}%
\providecommand \enquote  [1]{``#1''}%
\providecommand \bibnamefont  [1]{#1}%
\providecommand \bibfnamefont [1]{#1}%
\providecommand \citenamefont [1]{#1}%
\providecommand \href@noop [0]{\@secondoftwo}%
\providecommand \href [0]{\begingroup \@sanitize@url \@href}%
\providecommand \@href[1]{\@@startlink{#1}\@@href}%
\providecommand \@@href[1]{\endgroup#1\@@endlink}%
\providecommand \@sanitize@url [0]{\catcode `\\12\catcode `\$12\catcode
  `\&12\catcode `\#12\catcode `\^12\catcode `\_12\catcode `\%12\relax}%
\providecommand \@@startlink[1]{}%
\providecommand \@@endlink[0]{}%
\providecommand \url  [0]{\begingroup\@sanitize@url \@url }%
\providecommand \@url [1]{\endgroup\@href {#1}{\urlprefix }}%
\providecommand \urlprefix  [0]{URL }%
\providecommand \Eprint [0]{\href }%
\@ifxundefined \urlstyle {%
  \providecommand \doi  [0]{\begingroup \@sanitize@url \@doi}%
  \providecommand \@doi [1]{\endgroup \@@startlink {\doibase
  #1}doi:\discretionary {}{}{}#1\@@endlink }%
}{%
  \providecommand \doi  [0]{doi:\discretionary{}{}{}\begingroup
  \urlstyle{rm}\Url }%
}%
\providecommand \doibase [0]{http://dx.doi.org/}%
\providecommand \Doi [0]{\begingroup \@sanitize@url \@Doi }%
\providecommand \@Doi  [1]{\endgroup\@@startlink{\doibase#1}\@@Doi}%
\providecommand \@@Doi [1]{#1\@@endlink}%
\providecommand \selectlanguage [0]{\@gobble}%
\providecommand \bibinfo  [0]{\@secondoftwo}%
\providecommand \bibfield  [0]{\@secondoftwo}%
\providecommand \translation [1]{[#1]}%
\providecommand \BibitemOpen [0]{}%
\providecommand \bibitemStop [0]{}%
\providecommand \bibitemNoStop [0]{.\EOS\space}%
\providecommand \EOS [0]{\spacefactor3000\relax}%
\providecommand \BibitemShut  [1]{\csname bibitem#1\endcsname}%
\bibitem [{\citenamefont {Shannon}(1948)}]{Shannon:1948fk}%
  \BibitemOpen
  \bibfield  {author} {\bibinfo {author} {\bibfnamefont {C.~E.}\ \bibnamefont
  {Shannon}},\ }\href@noop {} {\bibfield  {journal} {\bibinfo  {journal} {The
  Bell System Technical Journal},\ }\textbf {\bibinfo {volume} {XXVII}},\
  \bibinfo {pages} {379} (\bibinfo {year} {1948})}\BibitemShut {NoStop}%
\bibitem [{\citenamefont {Rovelli}(1996)}]{Rovelli:1995fv}%
  \BibitemOpen
  \bibfield  {author} {\bibinfo {author} {\bibfnamefont {C.}~\bibnamefont
  {Rovelli}},\ }\href@noop {} {\bibfield  {journal} {\bibinfo  {journal} {Int.
  J. Theor. Phys.},\ }\textbf {\bibinfo {volume} {35}},\ \bibinfo {pages}
  {1637} (\bibinfo {year} {1996})},\ \Eprint {http://arxiv.org/abs/9609002}
  {arXiv:9609002 [quant-ph]} \BibitemShut {NoStop}%
\bibitem [{\citenamefont {Bronstein}(1936){\natexlab{a}}}]{Bronstein:1936kx}%
  \BibitemOpen
  \bibfield  {author} {\bibinfo {author} {\bibfnamefont {M.~P.}\ \bibnamefont
  {Bronstein}},\ }\href@noop {} {\bibfield  {journal} {\bibinfo  {journal} {Zh.
  Eksp. Tear. Fiz.},\ }\textbf {\bibinfo {volume} {6}},\ \bibinfo {pages} {195}
  (\bibinfo {year} {1936}{\natexlab{a}})}\BibitemShut {NoStop}%
\bibitem [{\citenamefont {Bronstein}(1936){\natexlab{b}}}]{Bronstein:9vn}%
  \BibitemOpen
  \bibfield  {author} {\bibinfo {author} {\bibfnamefont {M.~P.}\ \bibnamefont
  {Bronstein}},\ }\href@noop {} {\bibfield  {journal} {\bibinfo  {journal}
  {Phys. Z. Sowjetunion},\ }\textbf {\bibinfo {volume} {9}},\ \bibinfo {pages}
  {140} (\bibinfo {year} {1936}{\natexlab{b}})}\BibitemShut {NoStop}%
\bibitem [{\citenamefont {Rovelli}\ and\ \citenamefont
  {Smolin}(1995)}]{Rovelli:1994ge}%
  \BibitemOpen
  \bibfield  {author} {\bibinfo {author} {\bibfnamefont {C.}~\bibnamefont
  {Rovelli}}\ and\ \bibinfo {author} {\bibfnamefont {L.}~\bibnamefont
  {Smolin}},\ }\href@noop {} {\bibfield  {journal} {\bibinfo  {journal} {Nucl.
  Phys.},\ }\textbf {\bibinfo {volume} {B442}},\ \bibinfo {pages} {593}
  (\bibinfo {year} {1995})},\ \Eprint {http://arxiv.org/abs/9411005}
  {arXiv:9411005 [gr-qc]} \BibitemShut {NoStop}%
\bibitem [{\citenamefont {Ashtekar}\ and\ \citenamefont
  {Lewandowski}(1997)}]{Ashtekar:1996eg}%
  \BibitemOpen
  \bibfield  {author} {\bibinfo {author} {\bibfnamefont {A.}~\bibnamefont
  {Ashtekar}}\ and\ \bibinfo {author} {\bibfnamefont {J.}~\bibnamefont
  {Lewandowski}},\ }\href@noop {} {\bibfield  {journal} {\bibinfo  {journal}
  {Class. Quant. Grav.},\ }\textbf {\bibinfo {volume} {14}},\ \bibinfo {pages}
  {A55} (\bibinfo {year} {1997})},\ \Eprint {http://arxiv.org/abs/9602046}
  {arXiv:9602046 [gr-qc]} \BibitemShut {NoStop}%
\end{thebibliography}
 \end{document}